\documentclass[%
 reprint,
 amsmath,amssymb,
 aps,
]{revtex4-2}
\usepackage[T1]{fontenc}
\usepackage{amsmath}
\usepackage{graphicx}% Include figure files
\usepackage{dcolumn}% Align table columns on decimal point
\usepackage{bm}% bold math
\usepackage{footnote}
\usepackage{subfigure}% 
\usepackage{hyperref}% add hypertext capabilities
\hypersetup{colorlinks=true, citecolor=blue, urlcolor=blue, linkcolor=blue}
\begin{document}

\title{Bose-Einstein Condensation of Three-Dimensional Exciton-Polaritons}

\author{Junhui Cao$^{1}$}%
\email{tsao.c@mipt.ru}
\author{Alexey Kavokin$^{1,2,3}$}%
\email{a.kavokin@westlake.edu.cn}
\affiliation{$^{1}$Abrikosov Center for Theoretical Physics, Moscow Center for Advanced Studies, Moscow 141701, Russia\\
$^{2}$School of Science, Westlake University, 18 Shilongshan Road, Hangzhou 310024, Zhejiang Province, China\\
$^{3}$Department of Physics, St. Petersburg State University, University Embankment, 7/9, St. Petersburg, 199034, Russia}

\begin{abstract}
We develop a band-structure-based theory of exciton-polaritons in a three-dimensional inverse-opal photonic crystal doped with semiconductor quantum dots. Starting from a symmetry-selected bright photonic branch near the photonic gap edge, we construct an exciton-photon Hamiltonian and obtain a lower-polariton band with a W-point global minimum and a nearby X-point van-Hove-enhanced density of states. We show that the W valleys determine the equilibrium Bose-Einstein condensation threshold, while the X-point saddle provides a finite excited-state capacity that renormalizes the critical temperature when the W-X offset is thermally accessible. By tuning the exciton resonance and the light-matter coupling, the relative W-X ordering can be reconstructed, leading to a strong variation of the critical temperature. We further formulate a momentum-resolved Boltzmann model for driven-dissipative kinetics. Under non-resonant pumping, reservoir feeding, radiative decay, and inter-sector relaxation can produce either W-dominated condensation, a mixed W-X regime, or an X-dominated nonequilibrium coherent state. Our results establish three-dimensional photonic-crystal polaritons as a platform where condensation is controlled not only by the band minimum but also by valley geometry, van-Hove-enhanced phase space, and relaxation pathways.
\end{abstract}

\maketitle
\section{Introduction}
\label{sec:intro}

Exciton-polaritons are hybrid light-matter quasiparticles formed by the strong coupling between confined electromagnetic modes and excitonic resonances \cite{weisbuch1992observation,cao2021strong,butov2012behaviour,kavokin2017microcavities,wouters2007excitations}. Their small effective mass, inherited from the photonic component, makes them a universal platform for collective bosonic phenomena, while their excitonic component provides interactions and nonlinearities \cite{richard2005experimental,deng2010exciton,kasprzak2006bose,wouters2008spatial}. Most studies of polariton condensation have so far been carried out in planar or other one-/two-dimensional structures, where the low-energy dispersion is typically governed by a single cavity mode or a structured lattice band, and the condensation is typically discussed either in the context of Berezinskii-Kosterlitz-Thouless physics or in the broader driven-dissipative setting of nonequilibrium polariton condensation \cite{imamog1996nonequilibrium,hu2017imaging,nitsche2014algebraic,klembt2018exciton,whittaker2018exciton,cao2026emergent}. By contrast, a three-dimensional photonic crystal offers a qualitatively different setting, where the polariton may inherit a three-dimensional Bloch band structure, with symmetry-enforced degeneracies, anisotropic effective masses, and nontrivial density-of-states geometry near the band edge \cite{lin1998three,moon2010chemical,cersonsky2021diversity,fleming2002all}. Besides, a three-dimensional polaritonic band with a quadratic low-energy minimum can support a true Bose-Einstein condensation transition due to a converged infrared statistical physics of the condensate. In a three-dimensional Bloch band, the condensation problem is controlled by the location of the global minimum and the structure of nearby saddle points and associated van Hove-enhanced regions of phase space \cite{pitaevskii2016bose,ibanescu2006enhanced}. The periodic 3D photonic environment reshapes the radiative continuum and can substantially suppress photonic losses near symmetry-selected band-edge modes, which is beneficial for realizing the strong-coupling of light and excitons \cite{aoki2008coupling,tandaechanurat2011lasing}. Moreover, in a photonic crystal, not every photonic mode near the gap edge couples equally to the exciton. The relevant low-energy polariton branch must be selected jointly by momentum matching, little group symmetry, polarization compatibility, and spatial field overlap. As a result, before one can address condensation, one must first identify the symmetry-selected bright photonic branch that actually participates in the light-matter hybridization.

The present work is motivated by this band-structure-driven viewpoint. We consider a three-dimensional face-centered cubic (FCC) photonic crystal whose exciton-polariton lower branch exhibits two key features: a global minimum at the W-point and a nearby saddle-point-enhanced density of states near the X-point \cite{miguez1998control,doosje2000photonic,tarhan1996photonic}. This combination makes the system particularly interesting. On the one hand, the W-point minimum controls the low-energy thermodynamics and provides the natural candidate for equilibrium Bose-Einstein condensation. On the other hand, the nearby X-point high-density-of-states region provides a competing channel for population accumulation under pumping and incomplete relaxation.

The first objective of this study is to construct a symmetry-based theory of three-dimensional exciton-polaritons near the photonic gap edge. Starting from the numerically obtained photonic band structure of the FCC inverse opal photonic crystal, we retain only the dominant bright photonic branch selected by symmetry and field overlap, and couple it to a weakly dispersive exciton mode. This yields the light-matter interaction Hamiltonian that determines the lower- and upper-polariton branches, their Hopfield coefficients, and the low-energy curvature near the polariton minimum. The second objective is to determine how a nearby van Hove-enhanced density of states modifies Bose-Einstein condensation in three dimensions. For a generic three-dimensional quadratic minimum, the low-energy density of states scales as $\rho(E)\propto \sqrt{E-E_0}$, which allows a finite equilibrium condensation threshold. However, if an enhanced density of states is located only a small energy above the global minimum, the excited-state can store more particles than in a purely ideal parabolic band. The condensate still forms at the global minimum in equilibrium, but the critical density and critical temperature are renormalized by the nearby band-structure feature. Finally, we address the driven-dissipative kinetic problem. Exciton-polaritons are open-dissipative quasiparticles sustained by pumping and limited by loss, hence the experimentally realized occupation pattern is determined by the competition between dispersion, injection, decay, and momentum-space relaxation. In this regime, a high-density-of-states region near X-point can be preferentially populated under momentum-selective pumping even though the true lower-polariton minimum lies at W-point. Understanding this competition requires a kinetic description that goes beyond equilibrium thermodynamics and explicitly tracks the redistribution of population across momentum space.

The paper is organized as follows. In Sec.~\ref{sec:symmetry_model}, we construct the symmetry-selected model for three-dimensional exciton-polaritons and derive the lower-polariton dispersion near the photonic gap edge. In Sec.~\ref{sec:vH_BEC}, we analyze the effect of a nearby van Hove-enhanced density of states on the equilibrium-like Bose-Einstein condensation threshold. In Sec.~\ref{sec:regime} and \ref{sec:kinetics}, we formulate a Boltzmann kinetic description for momentum-selective condensation and study the competition between pumping into the X-point high-density-of-states region and relaxation toward the W-point global minimum. In Sec. \ref{sec:expe}, we discuss the possible experimental signature and summarize the main physical conclusions and discuss their implications for three-dimensional photonic-crystal polaritons.

\begin{figure}
    \centering
    \includegraphics[width=1\linewidth]{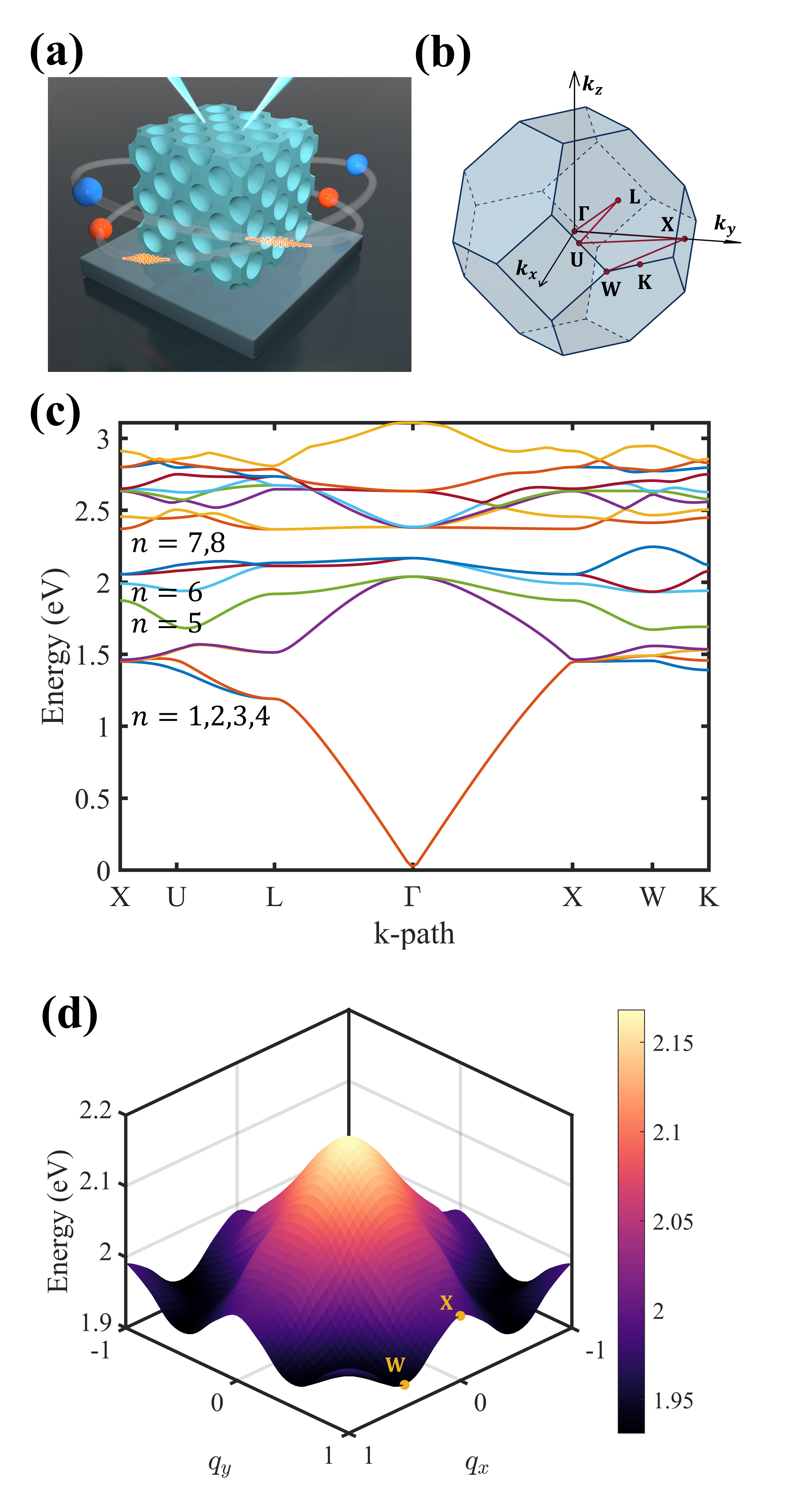}
    \caption{(a) Scheme of the inverse opal FCC photonic crystal. (b) First Brillouin zone and high symmetry path. (c) Band structure along the high-symmetry path. (d) Band structure of $n=6$ in $(k_x,k_y)$ plane at $k_z=0$, where $q_{x,y}=\frac{a}{2\pi}k_{x,y}$ is the normalized wavevector.}
    \label{fig:intro_band}
\end{figure}

\section{The band structure of a three-dimensional polariton crystal}
\label{sec:symmetry_model}

We begin by constructing a low-energy description of exciton-polaritons in a three-dimensional photonic crystal. The central point of this section is that, near the photonic gap edge, the relevant polariton branch cannot be identified solely by energetic proximity. Instead, it must be selected jointly by Bloch momentum conservation, crystal symmetry, polarization compatibility, and spatial field overlap with the excitonic dipole \cite{hopfield1958theory,gerace2007quantum,zhang2018photonic,whittaker2021exciton}. Once this symmetry-selected bright band is identified, one can formulate a reduced exciton-photon Hamiltonian that captures the lower-polariton dispersion, the location of its global minimum, and the nearby saddle-point structure that later controls both the equilibrium condensation threshold and the nonequilibrium kinetic flow.

Throughout this work, we consider a three-dimensional photonic crystal with face-centered-cubic symmetry, made of inverse opal structure of silicon (refractive index is set as 4, neglecting absorption) \cite{tetreault2004silicon,waterhouse2007opal,klimonsky2011photonic}. The lattice constant is 350 nm, with the diameter of each vacuum vacancy being 240 nm. Excitonic material has a resonance energy of 2.2 eV and is spread around the edge of vacancies. We denote the bare photonic Bloch bands by $E_{{\rm ph},n}(\mathbf{k})$, where $n$ labels the band index and $\mathbf{k}$ belongs to the first Brillouin zone. Our numerical analysis shows that the low-energy band structure of interest (band $n=6$) is organized around two high-symmetry regions: a global minimum at the W-point and a nearby saddle-point-enhanced structure at the X-point. More specifically, the target lowest photonic branch reaches its global minimum at W, where it forms a symmetry-enforced twofold degeneracy transforming as the $E$ irreducible representation of the little group $D_{2d}$. By contrast, the relevant saddle structure at X belongs to the $A_{2g}$ irreducible representation of $D_{4h}$ and is spectrally distinguishable from the nearby $E_g$ and $B_{1u}$ branches (band $n=7,8$ and $n=5$). These symmetry assignments provide the backbone of the effective theory.

\subsection{Bare photonic band and symmetry filtering}
\label{subsec:bare_photonic}

It is well-known that the strong coupling of an individual quantum dot and a photon mode in a microcavity is difficult to achieve, and the resulting vacuum field Rabi splitting might be small \cite{reithmaier2004strong,englund2010resonant}. Thousands of quantum dots embedded in a semiconductor microcavity would ensure much larger exciton-photon interaction strength, but inevitable inhomogeneous broadening of the excitonic resonance in an array of epitaxial quantum dots may strongly affect coherence of polariton modes that is significant for observation of quantum many-body effects \cite{vuckovic2023scalable}. In this context, embedding nearly identical colloidal quantum dots in empty spheres that compose a photonic crystal lattice of inverted opals may offer a good opportunity to achieve Rabi-splittings that exceed exciton and photon line broadenings, which is an essential requirement for the realization of strong exciton-photon coupling regime. In the present theoretical work we consider spatially homogeneous 3D polaritonic crystals, which is a helpful idealization enabling analytical calculation of important macroscopic characteristics of the polariton gas, such as the critical temperature of Bose-Einstein condensation. 

In the inverse-opal structure, the active excitonic medium is expected to be localized near the internal surface of the high-index dielectric skeleton, as the relevant band-edge Bloch fields are enhanced near the dielectric interface, and surface adsorption of quantum dots or emitters is energetically more favorable than a freely suspended pore-center configuration. The exciton-photon coupling is therefore viewed as an overlap with a wall-bound active region. Let $\mathbf{E}_{n\mathbf{k}}(\mathbf{r})$ denote the electric-field profile of the $n$-th photonic Bloch mode. The exciton-photon coupling at fixed Bloch momentum $\mathbf{k}$ is controlled microscopically by an overlap matrix element of the form
\begin{equation}
g_{n\alpha}(\mathbf{k})
=-\frac{1}{\hbar}
\int d^3r\,
\Phi^{*}_{X,\alpha}(\mathbf{r})\,
\mathbf{d}_{\alpha}\cdot \mathbf{E}_{n\mathbf{k}}(\mathbf{r}),
\label{eq:g_overlap}
\end{equation}
where $\Phi_{X,\alpha}(\mathbf{r})$ is the excitonic envelope associated with excitonic channel $\alpha$, and $\mathbf{d}_{\alpha}$ is the corresponding optical dipole matrix element, where the interaction is defined as $H_{\rm int}=-\mathbf{d}_{\alpha}\cdot \mathbf{E}_{n\mathbf{k}}(\mathbf{r})$. Equation \eqref{eq:g_overlap} implies that not every photonic mode near the band edge is relevant for the low-energy polariton problem. A given band contributes only if all three conditions are satisfied, namely the Bloch momentum conservation, a non-zero overlap, and a symmetry-allowed transition. These criteria motivate a symmetry filtering of the bare photonic spectrum. In the present system, this filtering leads to a distinguished bright low-energy photonic band continuously connected to the target lower polariton branch. Generically, away from symmetry-enforced degeneracies, this band can be represented by a single effective bright photonic mode. However, at special high-symmetry points such as W, the relevant Hilbert space is intrinsically multidimensional because the target band is symmetry-enforced to be twofold degenerate. It is therefore important to distinguish between a single bright band and a single scalar photonic degree of freedom: the former is the correct notion globally, whereas the latter is only valid locally away from symmetry-protected degeneracies.

This distinction is essential for the present problem. At the X-point, the relevant saddle structure is carried by an isolated $A_{2g}$ branch, so the local bright sector is one-dimensional. At the W-point, by contrast, the minimum belongs to a two-dimensional $E$ representation of $D_{2d}$, so the local bright sector is two-dimensional even before coupling to the exciton. The minimal low-energy theory must therefore be flexible enough to reduce to a $2\times2$ exciton-photon problem near generic momenta or near X, while retaining a larger local matrix structure near W.

%The disorder can be included in the model by adding a random stationary potential term in the generalized Gross-Pitaevskii equation for the polariton condensate. The analysis of disorder-induced effects such as spatial decoherence and pinning of vortices and solitons will be in the focus of our further studies.

\subsection{Exciton band and light-matter Hamiltonian}
\label{subsec:exciton_band}

We next specify the excitonic part of the theory. The exciton is modeled as a weakly dispersive bosonic mode,
\begin{equation}
E_X(\mathbf{k}) = E_X^0 + \frac{\hbar^2|\mathbf{k}|^2}{2M_X},
\label{eq:exc_disp}
\end{equation}
where $E_X^0$ is the bare exciton resonance and $M_X$ is the exciton mass. Since $M_X$ is typically much larger than the photonic effective mass scale, the exciton dispersion is weak over the momentum range relevant to the photonic band-edge structure. In most of the analysis below one may therefore regard the exciton as nearly flat, $E_X(\mathbf{k}) \approx E_X^0$, while retaining the small residual curvature only when needed for quantitative estimates. The symmetry-selected light-matter Hamiltonian may be written as
\begin{equation}
\begin{aligned}
\hat{H}
=
\sum_{\mathbf{k}}
\bigg [
&\sum_{\mu,\nu \in \mathcal{B}(\mathbf{k})}
\hat{a}^{\dagger}_{\mu\mathbf{k}}
H^{\rm ph}_{\mu\nu}(\mathbf{k})
\hat{a}_{\nu\mathbf{k}}
+E_X(\mathbf{k})\,\hat{b}^{\dagger}_{\mathbf{k}}\hat{b}_{\mathbf{k}}\\
&+
\sum_{\mu\in\mathcal{B}(\mathbf{k})}
\left(
g_{\mu}(\mathbf{k})\hat{a}^{\dagger}_{\mu\mathbf{k}}\hat{b}_{\mathbf{k}}
+
g_{\mu}^{*}(\mathbf{k})\hat{b}^{\dagger}_{\mathbf{k}}\hat{a}_{\mu\mathbf{k}}
\right)
\bigg ].
\end{aligned}
\label{eq:general_H}
\end{equation}
Here $\mathcal{B}(\mathbf{k})$ denotes the symmetry-selected bright photonic sector at momentum $\mathbf{k}$. Its dimension is one over most of the Brillouin zone, but becomes two near W because of the $E$ doublet. $a$ and $b$ denote the annihilation operators of photons and excitons, respectively.

\subsection{Generic single-channel polariton description}
\label{subsec:generic_2x2}

Away from special degeneracy points, the bright photonic Hilbert space is effectively one-dimensional, and the Hamiltonian reduces to the familiar $2\times2$ coupled-oscillator form
\begin{equation}
H_{\rm pol}(\mathbf{k})
=
\begin{pmatrix}
E_{\rm ph}(\mathbf{k}) & g(\mathbf{k})\\
g^{*}(\mathbf{k}) & E_X(\mathbf{k})
\end{pmatrix},
\label{eq:H_gen}
\end{equation}
where $E_{\rm ph}(\mathbf{k})$ is the selected bright photonic branch and $g(\mathbf{k})$ is the corresponding effective coupling strength.

The eigenvalues of Eq.~\eqref{eq:H_gen} define the upper- and lower-polariton branches,
\begin{equation}
\begin{aligned}
E_{\pm}(\mathbf{k})
=&
\frac{E_{\rm ph}(\mathbf{k})+E_X(\mathbf{k})}{2}\\
&\pm
\frac{1}{2}
\sqrt{
\left[E_{\rm ph}(\mathbf{k})-E_X(\mathbf{k})\right]^2
+
4|g(\mathbf{k})|^2
}.
\end{aligned}
\label{eq:polariton_branches}
\end{equation}
In the present work, the branch of primary interest is the lower polariton, $E_{LP}(\mathbf{k}) \equiv E_{-}(\mathbf{k})$, since it is this branch that hosts the global minimum relevant to condensation.

The corresponding eigenvectors determine the Hopfield coefficients ($C$ for photons and $X$ for excitons). Writing the lower-polariton annihilation operator as
\begin{equation}
\hat{p}_{\mathbf{k}}
=
C_{\mathbf{k}}\hat{a}_{\mathbf{k}}
+
X_{\mathbf{k}}\hat{b}_{\mathbf{k}},
\qquad
|C_{\mathbf{k}}|^2+|X_{\mathbf{k}}|^2=1,
\label{eq:hopfield_def}
\end{equation}
one finds
\begin{equation}
|X_{\mathbf{k}}|^2
=
\frac{1}{2}
\left[
1+
\frac{E_{\rm ph}(\mathbf{k})-E_X(\mathbf{k})}
{\sqrt{\left[E_{\rm ph}(\mathbf{k})-E_X(\mathbf{k})\right]^2+4|g(\mathbf{k})|^2}}
\right],
\label{eq:Xk}
\end{equation}
and
\begin{equation}
|C_{\mathbf{k}}|^2 = 1-|X_{\mathbf{k}}|^2.
\label{eq:Ck}
\end{equation}
These coefficients determine the local light-matter content of the lower branch and therefore control the effective mass, interaction scale, and lifetime. Also, they provide the natural momentum dependence of the decay rate entering the kinetic theory,
\begin{equation}
\gamma_{\mathbf{k}}
=
|C_{\mathbf{k}}|^2\gamma_{\rm ph}
+
|X_{\mathbf{k}}|^2\gamma_X,
\label{eq:gamma_k}
\end{equation}
where $\gamma_{\rm ph}$ and $\gamma_X$ are the bare photonic and excitonic loss rates.

\subsection{Local effective theory near the W-point minimum}
\label{subsec:W_local}

The global minimum of the target photonic branch occurs at the W-point. At this momentum, the relevant bare photonic band transforms as the two-dimensional irreducible representation $E$ of the little group $D_{2d}$. Let $\mathbf{W}$ denote the corresponding Bloch momentum and define $\mathbf{q}=\mathbf{k}-\mathbf{W}$. In a local basis spanning the photonic $E$ doublet, the bare photonic Hamiltonian takes the general form
\begin{equation}
H^{\rm ph}_{W}(\mathbf{q})
=
\omega^{\rm ph}_{W}\,\mathbf{I}_{2}
+
d_0(\mathbf{q})\,\mathbf{I}_{2}
+
d_x(\mathbf{q})\,\sigma_x
+
d_y(\mathbf{q})\,\sigma_y
+
d_z(\mathbf{q})\,\sigma_z,
\label{eq:Hph_W}
\end{equation}
where $\sigma_{x,y,z}$ are Pauli matrices acting in the local doublet space and the functions $d_{\mu}(\mathbf{q})$ are constrained by $D_{2d}$ symmetry. At $\mathbf{q}=0$, the two bare photonic modes are degenerate, so $\mathbf{d}(\mathbf{0})=\mathbf{0}$.
The scalar term $d_0(\mathbf{q})$ governs the average curvature of the doublet, while the vector $\left(d_x,d_y,d_z\right)$ encodes symmetry-allowed splittings away from W. If one retains a single bright exciton branch, the minimal local light-matter Hamiltonian near W becomes
\begin{equation}
H_{W}(\mathbf{q})
=
\begin{pmatrix}
H^{\rm ph}_{W}(\mathbf{q}) & \mathbf{g}_{W}(\mathbf{q})\\
\mathbf{g}^{\dagger}_{W}(\mathbf{q}) & E_X(\mathbf{W}+\mathbf{q})
\end{pmatrix},\ \ 
\mathbf{g}_{W}(\mathbf{q})=
\begin{pmatrix}
g_{1}(\mathbf{q})\\
g_{2}(\mathbf{q})
\end{pmatrix}.
\label{eq:HW_local}
\end{equation}
Equation \eqref{eq:HW_local} is a $3\times3$ matrix and represents the local model compatible with a photonic $E$ doublet from band $6,7$, and a single excitonic branch.

The precise leading-order form of $\mathbf{g}_W(\mathbf{q})$ depends on how the excitonic channel transforms under the local symmetry and on the detailed spatial embedding of the active excitonic medium. In a perfectly symmetric setting, a scalar excitonic channel may fail to couple linearly to the full $E$ doublet at $\mathbf{q}=0$, in which case the leading coupling appears at higher order in $\mathbf{q}$ or through a reduced-symmetry embedding. Second, if the excitonic band itself contains two symmetry-related bright channels near W, one may extend Eq.~\eqref{eq:HW_local} to a $4\times4$ model in which both the photonic and excitonic subspaces are two-dimensional. The present analysis does not require that level of generality, but the possibility is conceptually important.

For the purposes of the present paper, the crucial output of the local W-point theory is not the detailed matrix structure of every allowed term, but the identification of the low-energy quantities carried by the lower polariton near the minimum $E_0$. Let $E_{LP}(\mathbf{k})$ denote the lowest eigenvalue continuously connected to the target lower branch. Expanding around the minimum gives
\begin{equation}
E_{LP}(\mathbf{W}+\mathbf{q})
=
E_0
+
\frac{1}{2}
\sum_{ij}
A_{ij}q_i q_j
+
O(q^3),
\label{eq:LP_W_expand}
\end{equation}
where $E_0\equiv E_{LP}(\mathbf{W})$ and
\begin{equation}
A_{ij}
=
\left.
\frac{\partial^2 E_{LP}(\mathbf{k})}{\partial k_i \partial k_j}
\right|_{\mathbf{k}=\mathbf{W}}.
\label{eq:Aij_def}
\end{equation}
The corresponding inverse effective-mass tensor is
\begin{equation}
\left(M^{-1}_{LP}\right)_{ij}
=
\frac{1}{\hbar^2}A_{ij}.
\label{eq:mij}
\end{equation}
The symmetry-related multiplicity of equivalent W valleys in the Brillouin zone is denoted by $g_v$. This valley multiplicity does not alter the local dispersion in Eq.~\eqref{eq:LP_W_expand}, but it does enter the total low-energy density of states and the counting of low-energy occupation channels. The anisotropic curvature of the three-dimensional lower-polariton valley also has direct implications for possible superfluid transport, where the Bogoliubov sound velocity becomes direction dependent,
\[
c(\hat{\mathbf{q}})=
\sqrt{
g_{\rm LP}n_0\,
\hat{\mathbf{q}}^{T}M_{\rm LP}^{-1}\hat{\mathbf{q}}
}.
\]
Consequently, both the Landau critical velocity and the superfluid stiffness are expected to be anisotropic in the three-dimensional Bloch band. A quantitative treatment of the interacting and driven-dissipative superfluid response is left for future work.

\subsection{Local theory near the X-point saddle and van Hove scale}
\label{subsec:X_local}

We next consider the X-point structure, which is central to both the density-of-states enhancement and the nonequilibrium kinetic competition discussed later. Let $\mathbf{X}$ denote the relevant X-point momentum and define $\mathbf{q}=\mathbf{k}-\mathbf{X}$. In the present system, the target saddle structure at X belongs to the one-dimensional irreducible representation $A_{2g}$ of $D_{4h}$ and is spectrally distinguishable from the nearby $E_g$ and $B_{1u}$ branches. This separation makes X a particularly clean point at which the generic single-channel description applies.

The local Hamiltonian near X is therefore
\begin{equation}
H_{X}(\mathbf{q})
=
\begin{pmatrix}
E^{\rm ph}_{A_{2g}}(\mathbf{X}+\mathbf{q}) & g_{X}(\mathbf{q})\\
g^{*}_{X}(\mathbf{q}) & E_X(\mathbf{X}+\mathbf{q})
\end{pmatrix},
\label{eq:HX_local}
\end{equation}
with
\begin{equation}
E^{\rm ph}_{A_{2g}}(\mathbf{X}+\mathbf{q})
=
\omega^{\rm ph}_{X}
+
\alpha_1 q_1^2
-
\alpha_2 q_2^2
+
\alpha_3 q_3^2
+
\cdots.
\label{eq:saddle_ph}
\end{equation}
Here $q_{1,2,3}$ are local momentum coordinates chosen so that the saddle geometry is explicit. The characteristic signature is the coexistence of positive and negative curvatures, which produces a saddle surface. After coupling to the exciton, the corresponding lower-polariton branch near X remains of the form
\begin{equation}
E_{LP}(\mathbf{X}+\mathbf{q})
=
E_{LP}(\mathbf{X})
+
\tilde{\alpha}_1 q_1^2
-
\tilde{\alpha}_2 q_2^2
+
\tilde{\alpha}_3 q_3^2
+
\cdots,
\label{eq:saddle_LP}
\end{equation}
with renormalized coefficients $\tilde{\alpha}_i$ determined by the hybridization.

The key quantity associated with the X-point saddle is the energy offset between the saddle structure and the global lower-polariton minimum,
\begin{equation}
\Delta_{XW}
=
E_{LP}(\mathbf{X})-E_{LP}(\mathbf{W}).
\label{eq:Delta_vH}
\end{equation}
This scale plays a dual role throughout the paper. In the equilibrium-like thermodynamic analysis, it controls how strongly the nearby van Hove-enhanced density of states modifies the condensation threshold. In the kinetic analysis, it sets the energetic distance over which particles injected near X must relax in order to reach the W-point minimum. It enhances the available phase space in a nearby excited-energy window. In the density of states this appears as a nontrivial enhancement centered at $E_0+\Delta_{XW}$. This distinction will be central in Sec.~\ref{sec:vH_BEC}.

\begin{figure}
    \centering
    \includegraphics[width=1\linewidth]{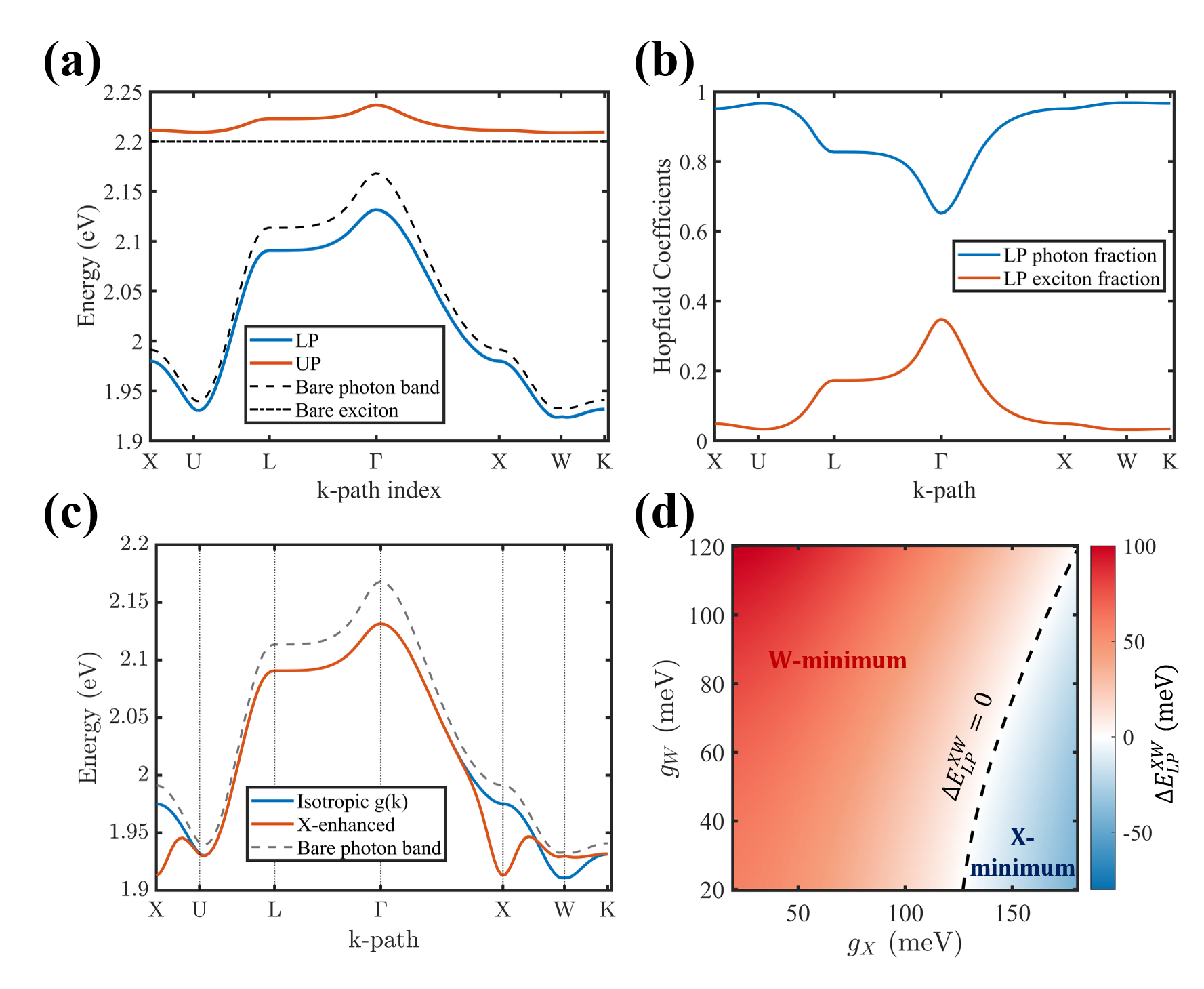}
\caption{
Lower-polariton band geometry and light-matter composition in the three-dimensional photonic crystal.
(a) Lower-polariton dispersion obtained from the symmetry-selected bright photonic branch coupled to a nearly flat exciton resonance. A uniform coupling constant $g=50$ meV is chosen.
(b) Corresponding Hopfield coefficient, showing the momentum-dependent excitonic/photon content of the lower-polariton branch.
(c) Lower-polariton dispersion in the X-minimum regime, showing that momentum-dependent light-matter coupling can reshape the local band geometry and turn the X-point into a minimum. Momentum-dependent coupling constants are $g_X=150$ meV, $g_W=30$ meV and background $g=50$ meV.
(d) Evolution of the relative W-X valley ordering as a function of the exciton resonance energy or detuning. The sign change of $\Delta_{XW}$ marks a Lifshitz-type reconstruction of the lowest-energy valley and separates the W-minimum and X-minimum regimes.
}

    \label{fig:lp_band}
\end{figure}

\section{Equilibrium Bose-Einstein Condensation of exciton-polaritons}
\label{sec:vH_BEC}

Having identified in Sec.~\ref{sec:symmetry_model} the symmetry-selected lower-polariton branch $E_{LP}(\mathbf{k})$, we now analyze its equilibrium-like condensation threshold. The key point is simple: the condensate is fixed by the global minimum at the symmetry-related W valleys, while the nearby X-point saddle enters only through the excited-state phase space of the same lower-polariton band.

Throughout this section we consider the near-equilibrium limit in which the polariton distribution is characterized by a temperature $T$ and a chemical potential $\mu<E_0$, where
\begin{equation}
E_0 = \min_{\mathbf{k}\in {\rm BZ}} E_{LP}(\mathbf{k}).
\label{eq:E0_sec2}
\end{equation}
The total density is then
\begin{equation}
n(T,\mu)
=
\int_{\rm BZ}\frac{d^3k}{(2\pi)^3}
\frac{1}{e^{\beta[E_{LP}(\mathbf{k})-\mu]}-1},
\qquad
\beta=\frac{1}{k_B T},
\label{eq:n_full_k}
\end{equation}
and equivalently, we have
\begin{equation}
n(T,\mu)
=
\int_{E_0}^{\infty} dE\,
\frac{\rho_{LP}(E)}{e^{\beta(E-\mu)}-1},
\label{eq:n_full_E}
\end{equation}
with
\begin{equation}
\rho_{LP}(E)
=
\int_{\rm BZ}\frac{d^3k}{(2\pi)^3}\,
\delta\!\left(E-E_{LP}(\mathbf{k})\right).
\label{eq:DOS_exact}
\end{equation}
At threshold, $\mu\to E_0^{-}$, so the maximum density that can be stored in excited states is
\begin{equation}
\begin{aligned}
n_c(T)
&=
\int_{\rm BZ}\frac{d^3k}{(2\pi)^3}
\frac{1}{e^{\beta[E_{LP}(\mathbf{k})-E_0]}-1}\\
&=
\int_{E_0}^{\infty} dE\,
\frac{\rho_{LP}(E)}{e^{\beta(E-E_0)}-1}.
\end{aligned}
\label{eq:nc_exact}
\end{equation}
This defines the critical temperature $T_c$ for a fixed total density $n$, with $n_c(T_c)=n$.

To isolate the universal low-energy part, we expand the lower branch around any one of the W valleys, $\mathbf{k}=\mathbf{W}_{\nu}+\mathbf{q}$,
\begin{equation}
E_{LP}(\mathbf{W}_{\nu}+\mathbf{q})
=
E_0+\frac{\hbar^2}{2}\,\mathbf{q}^{\rm T}M_W^{-1}\mathbf{q},
\label{eq:LP_W_quad}
\end{equation}
where $M_W$ is the effective-mass tensor and $g_v$ is the number of symmetry-related W valleys. The corresponding low-energy density of states is
\begin{equation}
\rho_W(E)
=
\frac{g_v}{4\pi^2}\left(\frac{2}{\hbar^2}\right)^{3/2}
\sqrt{\det M_W}\,\sqrt{E-E_0}\,\Theta(E-E_0),
\label{eq:rho_W}
\end{equation}
which gives the standard three-dimensional contribution
\begin{equation}
n_c^{W}(T)
=
\frac{g_v\,\zeta(3/2)}{(2\pi)^{3/2}\hbar^3}
\sqrt{\det M_W}\,(k_B T)^{3/2},
\label{eq:ncW_final}
\end{equation}
where $\zeta$ is the Riemann zeta function. Equation~\eqref{eq:ncW_final} shows that the W valleys support a universal three-dimensional Bose-Einstein-condensation threshold, where $n_c(T)\propto T^{3/2}$ and $T_c\propto n^{2/3}$, with the valley multiplicity entering only as a prefactor.

For practical calculations, the most useful representation is therefore the hybrid one:
\begin{equation}
n
=
\frac{g_v\,\zeta(3/2)}{(2\pi)^{3/2}\hbar^3}
\sqrt{\det M_W}\,(k_B T_c)^{3/2}
+\delta n_c(T_c),
\label{eq:Tc_hybrid_exact}
\end{equation}
where
\begin{equation}
\begin{aligned}
\delta n_c(T)&=n_c(T)-n_c^{W}(T),\\
&=\int_{E_0}^{\infty}dE\,\frac{\delta\rho(E)}{e^{(E-E_0)/(k_B T)}-1},
\end{aligned}
\end{equation}
is the numerically evaluated non-W contribution extracted from the full lower-polariton band. This form cleanly separates the universal W-valley physics from the finite-energy correction associated with the X-point DOS enhancement and the rest of the Brillouin zone. To make the role of the nearby X-point van Hove-enhanced region fully explicit, it is useful to decompose the critical density into contributions from different momentum-space sectors of the same lower-polariton band. The special role of the X-point sector is controlled by the energy offset $\Delta_{XW}$. Since $E_{LP}(X)$ is the characteristic energy of the X-point saddle region, the X-sector density of states has support only above $E_0+\Delta_{XW}$. The X-sector contribution to the excited-state capacity is
\begin{align}
n_c^{X}(T)
&=
\int_{E_0+\Delta_{XW}}^{\infty} dE\,
\frac{\rho_X(E)}
{e^{(E-E_0)/(k_B T)}-1}
\nonumber\\
&=
\int_{0}^{\infty} d\varepsilon\,
\frac{\bar{\rho}_X(\varepsilon)}
{e^{(\Delta_{XW}+\varepsilon)/(k_B T)}-1}.
\label{eq:ncX_exact}
\end{align}
Here $\rho_X(E)$ is the partial lower-polariton density of states associated with the momentum-space neighborhood of the X-saddle point. In three dimensions, this contribution is finite. The differentiation with respect to $\Delta_{XW}$ gives
\begin{align}
\frac{\partial n_c^{X}}{\partial \Delta_{XW}}
=
-\frac{1}{k_B T}
\int_{0}^{\infty} d\varepsilon\,
\bar{\rho}_X(\varepsilon)\,
\frac{e^{(\Delta_{XW}+\varepsilon)/(k_B T)}}
{\left[e^{(\Delta_{XW}+\varepsilon)/(k_B T)}-1\right]^2}.
\label{eq:dncXdDelta}
\end{align}
Since $\bar{\rho}_X(\varepsilon)\ge 0$, Eq.~\eqref{eq:dncXdDelta} implies $\frac{\partial n_c^{X}}{\partial \Delta_{XW}}<0$. Thus, increasing the X-point energy offset reduces the number of particles that can be stored at the X-point excited states, whereas lowering $\Delta_{XW}$ increases the contribution to the excited-state capacity of X-point.

\begin{figure}
    \centering
    \includegraphics[width=1\linewidth]{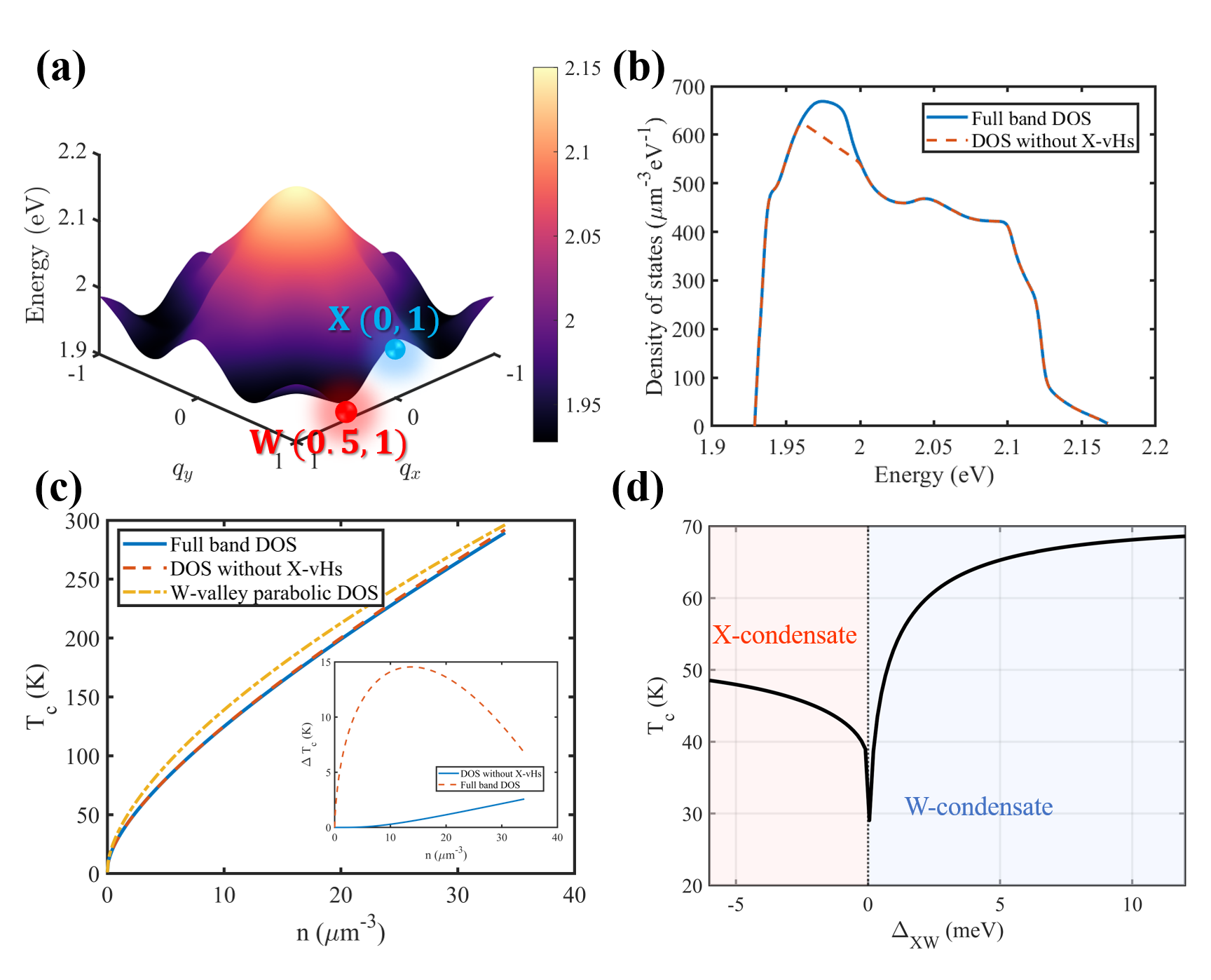}
\caption{
Density-of-states mechanism and equilibrium-like BEC threshold of the three-dimensional lower-polariton band.
(a) Band geometry of LP branch.
(b) Density of states. The full-band DOS (blue curve) is compared with a smooth reference DOS in which the X-sector contribution is removed (red dashed curve).
(c) Critical temperature as a function of total polariton density, calculated from the full lower-polariton DOS and compared with the W-valley parabolic approximation and the reference DOS without the X-sector contribution. The inset shows the difference of the critical temperature with corresponding corrections.
(d) Evolution of the critical temperature with the signed valley offset $\Delta_{XW}$.
}
    \label{fig:dos}
\end{figure}

Figure~\ref{fig:dos} summarizes the density-of-states mechanism and its consequence for the equilibrium-like BEC threshold. The key point is that the W and X regions play different thermodynamic roles when \(\Delta_{XW}>0\). The W valley is the minimum of the lower-polariton branch and therefore controls the infrared part of the Bose integral. This infrared behavior makes the excited-state capacity finite and leads to the standard three-dimensional scaling
$$
n_c^W(T)\propto T^{3/2},\qquad T_c\propto n^{2/3}.
$$
Therefore, the W valley fixes the equilibrium condensation energy and provides the natural low-energy channel for Bose-Einstein condensation. By contrast, the X-point saddle is located at the $E_{LP}(X)=E_0+\Delta_{XW}$. In three dimensions, this saddle appears as a finite excess contribution on top of the smooth lower-polariton background, as shown in Fig.~\ref{fig:dos}(b). Its influence on the equilibrium threshold is controlled by whether the offset \(\Delta_{XW}\) lies within the thermally occupied window of the Bose distribution. A reduction in \(\Delta_{XW}\) will increase the number of particles that can be stored in the X-sector.

This behavior is reflected in the critical-temperature curves in Fig.~\ref{fig:dos}(c). For the representative parameters considered here, the full-band result remains close to the reference curve without the X-sector contribution, indicating that the present X-point van Hove feature gives a finite correction to the critical temperature. The inset makes this correction explicit by showing the shift of \(T_c\) induced by the X-sector DOS. The effect becomes much stronger when the W-X valley offset is tuned. As shown in Fig.~\ref{fig:dos}(d), reducing \(\Delta_{XW}\) increases the X-sector contribution to the excited-state capacity and suppresses the critical temperature at fixed density. At $\Delta_{XW}=0$, the W and X valleys become degenerate and the low-energy phase space is maximally enlarged. When \(\Delta_{XW}\) changes sign, the global minimum switches from W to X, giving a Lifshitz-type reconstruction of the condensation valley \cite{lifshitz1960anomalies,volovik2017topological,jalali2023topological}.

\section{Relaxation Channels and Regime Classification}
\label{sec:regime}

The thermodynamic description above applies only when momentum-space redistribution is fast enough to establish a near-Bose distribution on the lower-polariton band before particles are lost. In practice, the same band geometry also supports a nonequilibrium channel: particles can be injected preferentially into the high-density-of-states X region and may or may not relax efficiently to the W valleys.

We use a reduced kinetic description in which the lower-polariton branch is projected onto two momentum-space sectors: the W-valley sector and the X-saddle sector. Their occupations are denoted by \(N_W\) and \(N_X\), and the relative dominance of the W sector is measured by
\begin{equation}
f_W=\frac{N_W}{N_W+N_X}.
\end{equation}

The two most important kinetic parameters are the effective feeding selectivity and the inter-sector relaxation rate. We define $S_{X/W}$ as the ratio of effective pumping intensities to X and W sectors, and measure the redistribution from X to W by the dimensionless ratio
\begin{equation}
r_{XW}=\frac{\Gamma_{XW}}{\gamma_X},
\end{equation}
where \(\Gamma_{XW}\) is the X$\to$W transfer rate and \(\gamma_X\) is the decay rate of the X-sector polaritons. The upward transfer rate \(\Gamma_{WX}\) may be included through a detailed-balance factor when a near-thermal bath is assumed, but the main competition is already captured by \(S_{X/W}\) and \(r_{XW}\).

The physical meaning of \(S_{X/W}\) depends on the pumping scheme. For resonant or quasi-resonant pumping, particles are injected directly into a selected momentum region of the lower-polariton band. In this case,
\begin{equation}
S_{X/W}^{\rm res}=\frac{P_X}{P_W},
\end{equation}
where \(P_X\) and \(P_W\) are controlled by the pump center momentum and spectral width. Thus resonant pumping provides an external handle for selecting either the W valley or the X region.

For non-resonant pumping, the situation is different. The pump first creates a high-energy reservoir, and the reservoir subsequently feeds the lower-polariton band. The effective feeding rates are then determined by the overlap between the reservoir relaxation window, the sector-resolved density of states, and the excitonic Hopfield weight:
\begin{equation}
\Gamma_\alpha
=
\int_{\Omega_\alpha}dE\,
\rho_\alpha(E)|X_\alpha|^2 F(E),
\qquad
\alpha=X,W .
\end{equation}
And
\begin{equation}
F(E)
=
\exp\left[
-\frac{(E-E_{\rm inj})^2}{2\sigma_E^2}
\right],
\label{eq:pump}
\end{equation}
is the pump profile, where \(E_{\rm inj}\) is the characteristic injection energy and \(\sigma_E\) is the relaxation broadening. The corresponding non-resonant selectivity is
\begin{equation}
S_{X/W}^{\rm nr}=\frac{\Gamma_X}{\Gamma_W}.
\end{equation}
In this case the \(X/W\) feeding preference is not imposed directly by the incident momentum. It is generated by the band geometry and by the relaxation pathway from the reservoir into the lower-polariton branch, which will be shown in the next section.

% We therefore characterize the kinetics in terms of four quantities: a momentum-space relaxation time $\tau_{\rm rel}$, band-resolved loss times $\tau_{\rm loss}^{W}$ and $\tau_{\rm loss}^{X}$, an inter-region transfer time $\tau_{X\rightarrow W}$, and a pumping selectivity ratio
% \begin{equation}
% \mathcal{S}_{X/W}=\frac{\Pi_X}{\Pi_W},
% \label{eq:RXW_def}
% \end{equation}
% where $\Pi_W$ and $\Pi_X$ are the injection rates into the W and X regions, respectively.

% These quantities define three regimes. In the equilibrium-like regime,
% \begin{equation}
% \tau_{\rm rel}\ll \tau_{\rm loss}^{W},\qquad
% \tau_{\rm rel}\ll \tau_{\rm loss}^{X},\qquad
% \tau_{X\rightarrow W}\ll \tau_{\rm loss}^{X},
% \label{eq:eq_regime_compact}
% \end{equation}
% so particles injected near X relax to the W valleys before decaying. The condensation threshold is then governed by Sec.~\ref{sec:vH_BEC}. In the strongly nonequilibrium regime,
% \begin{equation}
% \mathcal{S}_{X/W}\gg 1,
% \qquad
% \tau_{X\rightarrow W}\gtrsim \tau_{\rm loss}^{X},
% \label{eq:noneq_regime_compact}
% \end{equation}
% so the X region acts as an active accumulation channel rather than as a passive excited-state correction. Between these limits lies a crossover regime in which relaxation, loss, and selective pumping all occur on comparable timescales. This is the regime where the nearby X-point van Hove structure leaves the strongest imprint on the observed occupation pattern.

\section{Boltzmann Kinetics of Momentum-Selective Condensation}
\label{sec:kinetics}

\subsection{Momentum-selective quasi-resonant pump}
To describe the driven-dissipative regime explicitly, we consider the lower-polariton occupation number $n_{\mathbf{k}}(t)$ and evolve it according to a momentum-resolved bosonic Boltzmann equation,
\begin{equation}
\begin{aligned}
\frac{\partial n_{\mathbf{k}}}{\partial t}
=&
P_{\mathbf{k}}
-\gamma_{\mathbf{k}}n_{\mathbf{k}}+\int_{\rm BZ}\frac{d^3k'}{(2\pi)^3}
\Big [
W_{\mathbf{k}'\rightarrow\mathbf{k}}n_{\mathbf{k}'}(1+n_{\mathbf{k}})\\
&-W_{\mathbf{k}\rightarrow\mathbf{k}'}n_{\mathbf{k}}(1+n_{\mathbf{k}'})
\Big ].
\end{aligned}
\label{eq:boltzmann_full}
\end{equation}
Here $P_{\mathbf{k}}$ is the incoherent injection profile, $\gamma_{\mathbf{k}}$ is the mode-dependent decay rate, and $W_{\mathbf{k}\rightarrow\mathbf{k}'}$ is a redistribution kernel satisfying detailed balance in the absence of pumping and loss. The microscopic origin of the redistribution kernel can be phonon-assisted relaxation, reservoir-mediated scattering, polariton-polariton collisions, or weak disorder-assisted transfer.

We model momentum-selective pumping by a normalized profile centered near a target momentum $\mathbf{k}_p$,
\begin{equation}
P_{\mathbf{k}}=P_0\,\mathcal{P}(\mathbf{k}),
\qquad
\int_{\rm BZ}\frac{d^3k}{(2\pi)^3}\,\mathcal{P}(\mathbf{k})=1,
\label{eq:P_k_def}
\end{equation}
with $\mathbf{k}_p\approx \mathbf{X}$ for X-selective injection, $P_0$ the injection rate. The decay rate is inherited from the Hopfield composition of the lower branch,
\begin{equation}
\gamma_{\mathbf{k}}
=
|C_{\mathbf{k}}|^2\gamma_{\rm ph}+|X_{\mathbf{k}}|^2\gamma_X,
\label{eq:gamma_k_section4}
\end{equation}
so both the lifetime and the relaxation pathways are tied directly to the same band structure discussed in Sec.~\ref{sec:symmetry_model}.

It is useful to project the dynamics onto three momentum-space sectors,
\begin{equation}
{\rm BZ}=\Omega_W\cup\Omega_X\cup\Omega_R,
\label{eq:BZ_partition_section4}
\end{equation}
corresponding to the W valleys, the X-point high-density-of-states region, and the remainder of the band. Defining the average occupation
\begin{equation}
N_\alpha(t)=\frac{1}{\mathcal{D}_\alpha}
\int_{\Omega_\alpha}\frac{d^3k}{(2\pi)^3}n_k(t),
\qquad
\mathcal{D}_\alpha=
\int_{\Omega_\alpha}\frac{d^3k}{(2\pi)^3},
\label{eq:NW_NR_NX}
\end{equation}
where $\alpha\in\{W,X,R\}$, one obtains a reduced kinetic picture in which the physically relevant competition is transparent: direct pumping can strongly populate X, while redistribution transfers population downhill toward W. The corresponding inter-region flux is
\begin{equation}
\Phi_{X\rightarrow W}[n]
=
\int_{\Omega_X}\frac{d^3k}{(2\pi)^3}
\int_{\Omega_W}\frac{d^3k'}{(2\pi)^3}
W_{\mathbf{k}\rightarrow\mathbf{k}'}\,n_{\mathbf{k}}(1+n_{\mathbf{k}'}),
\label{eq:PhiXW_compact}
\end{equation}
and the associated transfer time is approximately $\tau_{X\to W}\approx \mathcal D_X N_X/\Phi_{X\to W}$.

This formulation leads to three generic outcomes. When redistribution dominates over loss, the steady state approaches a near-Bose distribution and the instability is W-dominated, recovering the equilibrium-like threshold problem of Sec.~\ref{sec:vH_BEC}. When X-selective pumping is strong and the downhill transfer is inefficient, the X region becomes a kinetic bottleneck and can host the dominant pre-threshold accumulation. In the crossover regime, both W and X carry appreciable weight, and the observed threshold depends on the full coupled solution of Eq.~\eqref{eq:boltzmann_full}.

In practice, the threshold can be diagnosed from the steady state reached under increasing pump power $P_0$. The central qualitative question is not whether W remains the true single-particle minimum, but whether the instability that first develops under pumping is weighted mainly in the W-valleys or in the X-region. This distinction is the nonequilibrium counterpart of the thermodynamic separation made in Sec.~\ref{sec:vH_BEC}, where W determines the true equilibrium condensate, while X reshapes the pathway by which the driven system reaches or avoids that condensate.

\subsection{Non-resonant Pump}
Here we consider the case of non-resonant optical excitation. In this situation, polaritons are not injected directly into a well-defined momentum state of the lower-polariton branch. Instead, the pump creates a high-energy excitonic reservoir, which subsequently feeds the lower-polariton band through phonon-assisted, disorder-assisted, or exciton-mediated relaxation processes.

We introduce a reservoir population \(N_R\) and write
\begin{equation}
P_{\mathbf{k}}^{\rm nr}
=
R_{\mathbf{k}} N_R ,
\end{equation}
where \(R_{\mathbf{k}}\) is the scattering rate from the reservoir into the lower-polariton state \(\mathbf{k}\). The reservoir dynamics is described by
\begin{equation}
\frac{dN_R}{dt}
=
P_0-\gamma_R N_R
-
N_R\int_{\rm BZ}\frac{d^3k}{(2\pi)^3}
R_{\mathbf{k}}\left(1+n_{\mathbf{k}}\right),
\end{equation}
where \(P_0\) is the non-resonant pump power and \(\gamma_R\) is the reservoir decay rate. For an excitonic reservoir, the scattering rate into the lower-polariton branch is expected to scale with the excitonic Hopfield coefficient. We therefore use the phenomenological form
\begin{equation}
R_{\mathbf{k}}
=
R_0 |X_{\mathbf{k}}|^2
F\!\left[E_{\rm LP}(\mathbf{k})\right],
\end{equation}
where \(R_0\) is scattering rate of reservoir to polariton. The function \(F(E)\) describes the energetic relaxation window from the reservoir to the lower-polariton band. We use the same Gaussian pump $F(E)$ as shown in Eq.~\eqref{eq:pump}.

Projecting this reservoir-fed source onto the momentum-space sectors \(\Omega_W\), \(\Omega_X\), and \(\Omega_R\), one obtains
\begin{equation}
P_\alpha^{\rm nr}
=
N_R\Gamma_\alpha ,
\qquad
\Gamma_\alpha
=
\int_{\Omega_\alpha}
\frac{d^3k}{(2\pi)^3}
R_{\mathbf{k}},
\qquad
\alpha=W,X,R.
\end{equation}
Unlike in resonant pumping, the feeding selectivity $S_{X/W}^{\rm nr}=\frac{\Gamma_X}{\Gamma_W}$ is generated by the combined effect of the lower-polariton density of states, the excitonic Hopfield weight, and the energetic relaxation window. As a result, the X-point saddle region can be fed efficiently under non-resonant pumping even when the lower-polariton minimum lies at the W point.

\begin{figure}
    \centering
    \includegraphics[width=1\linewidth]{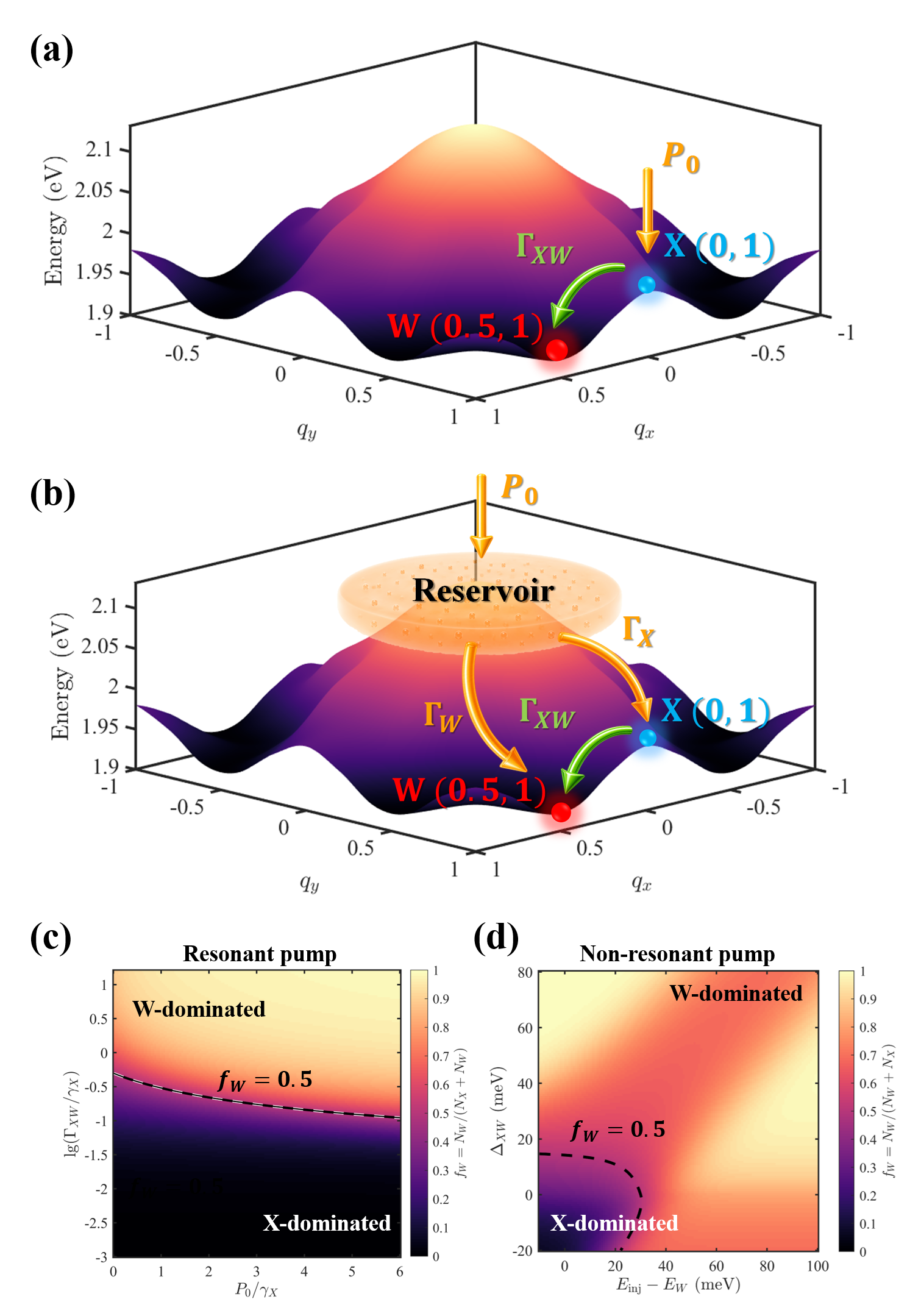}
    \caption{Dynamics for resonant pump and non-resonant pump. (a) Mechanism scheme for resonant pump. Pump injects polaritons into the X-point of the LP branch. (b) Mechanism scheme for non-resonant pump. (c) Population diagram for resonant pump. The white dashed curve indicates the analytical solution of $f_W=0.5$. (d) Population diagram for non-resonant pump.}
    \label{fig:dynamics}
\end{figure}

To visualize the kinetic competition, we simplify the full momentum-resolved Boltzmann equation to a sector-averaged model involving the reservoir, the X-point high-density-of-states region, and the W-valley minimum:
\begin{align}
\frac{dN_R}{dt}
&=
P_0-\gamma_R N_R
-\Gamma_X N_R(1+N_X)
-\Gamma_W N_R(1+N_W),
\\
\frac{dN_X}{dt}
&=
\Gamma_X N_R(1+N_X)
-\gamma_X N_X
-\Gamma_{XW}N_X(1+N_W),
\\
\frac{dN_W}{dt}
&=
\Gamma_W N_R(1+N_W)
-\gamma_W N_W
+\Gamma_{XW}N_X(1+N_W).
\end{align}
Here \(\gamma_X\) and \(\gamma_W\) are the sector-averaged polariton loss rates, while \(\Gamma_{XW}\) describes energy relaxation from the X region to the W valleys. The bosonic factors \(1+N_X\) and \(1+N_W\) account for final-state stimulation near threshold. When \(\Gamma_{XW}\gg \gamma_X\), particles fed into the X region relax to the W valleys before decaying, and the observed instability is W dominated. In contrast, when \(\Gamma_{XW}< \gamma_X\), the X region becomes a kinetic bottleneck. In this regime, the large density of states near the X-point saddle can lead to a strong X-centered population even though \(E_{\rm LP}(X)>E_{\rm LP}(W)\). Figure \ref{fig:dynamics} summarizes the resulting regime classification. For quasi-resonant excitation (a,c), the selectivity $S_{X/W}$ is externally controlled by the pump momentum and spectral width. The boundary $f_W=0.5$ separates the W-dominated and X-dominated steady states. The crossover boundary between the W-dominated and X-dominated regimes is estimated analytically by setting $f_W=1/2$ in the model. This gives the critical relaxation ratio:
\begin{equation}
\frac{\Gamma_{XW}^{c}}{\gamma_X}
=
\frac{\gamma_W/\gamma_X}
{1-\exp[-\Delta_{XW}/(k_B T_{\rm eff})]}
\frac{
1+\dfrac{P_0}{\gamma_X+\gamma_W}
}{
1+\dfrac{P_0}{N_{\rm sat}(\gamma_X+\gamma_W)}
}.  
\end{equation}
Here $\Delta_{XW}=E_{\rm LP}(X)-E_{\rm LP}(W)$, $T_{\rm eff}$ is the effective temperature controlling the W-to-X transfer, and $N_{\rm sat}$ characterizes the saturation of bosonic stimulation. We use $\gamma_X$ as the unit of rate and set $\gamma_W/\gamma_X=0.45$, $\Delta_{XW}=30$ meV, $T_{\rm eff}=150$ K, and $N_{\rm sat}=30$. The pump strength is varied in the range $P_0/\gamma_X=0.01$--$6$, while the relaxation ratio is scanned over $\Gamma_{XW}/\gamma_X=10^{-3}$--$10^{1.2}$. For non-resonant excitation (b,d), by contrast, the effective selectivity is generated internally by the lower-polariton density of states, the excitonic Hopfield weight, and the reservoir relaxation window. Therefore, an X-dominated population can emerge even when the true single-particle minimum remains at W, provided that the X-to-W relaxation rate is not large compared with the X-sector loss rate.

\section{Signatures and discussion}
\label{sec:expe}
The preceding analysis shows that the W and X regions play different roles in a three-dimensional polariton crystal. In thermal equilibrium, the condensate is selected by the global minimum of the lower-polariton branch. For the parameters considered here, this minimum is located at the symmetry-related W valleys. The X region, although higher in energy, remains important because it provides an enhanced finite-energy phase space associated with the saddle structure of the lower-polariton band. Under non-resonant pumping, this distinction becomes experimentally visible. The observed emission is determined by the competition between reservoir feeding, radiative loss, and relaxation from the X region toward the W valleys.

\begin{figure}
    \centering
    \includegraphics[width=1\linewidth]{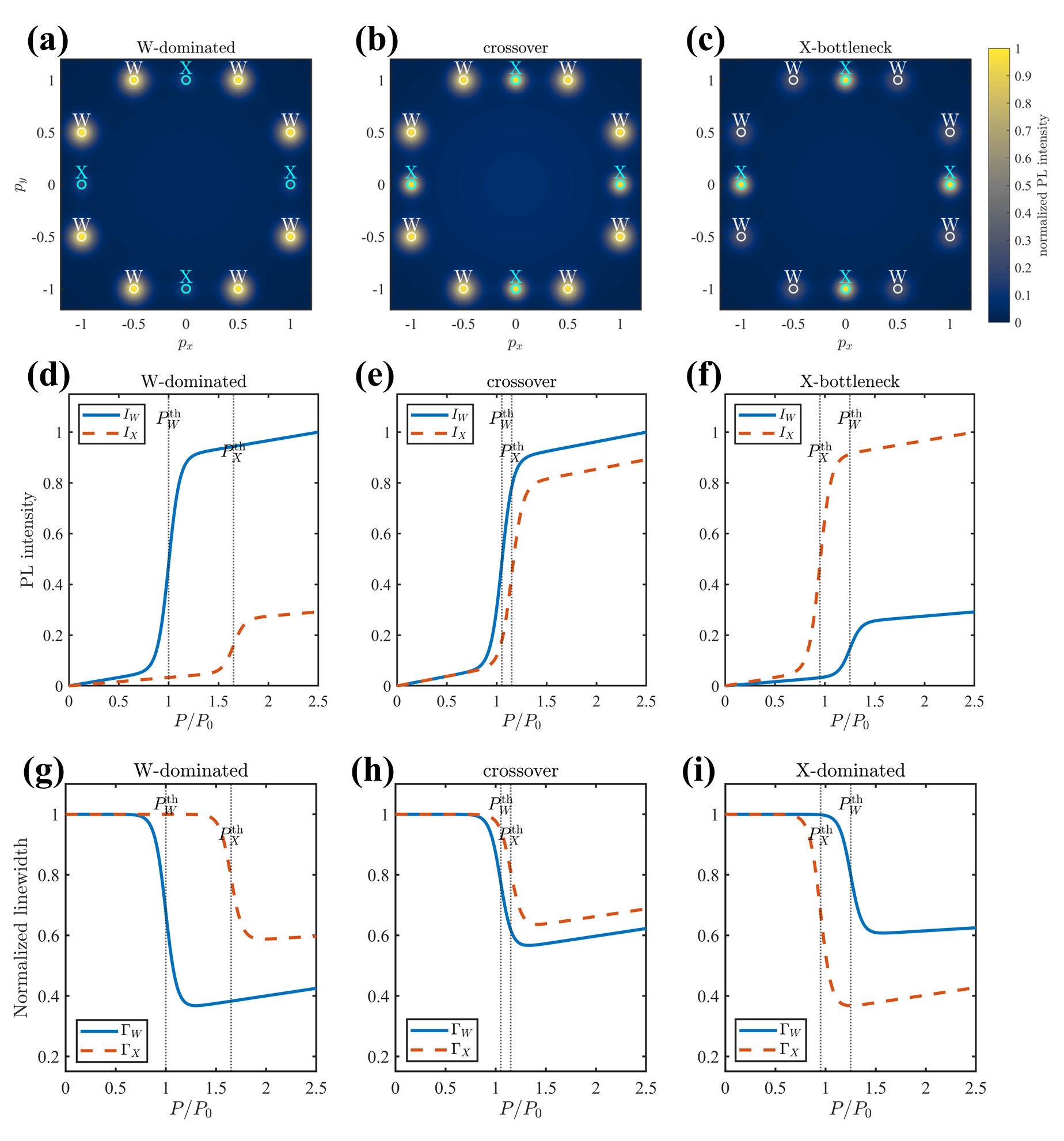}
    \caption{
    Signatures of W- and X-point polariton occupation.
    (a)-(c) Schematic momentum-resolved photoluminescence maps projected onto the \(k_z=0\) plane for the W-dominated, crossover, and X-dominated regimes. 
    The W-valleys and X-points are marked explicitly. 
    (d)-(f) Pump-power dependence of the momentum-integrated emission intensities \(I_W\) and \(I_X\). 
    The threshold-like increase occurs mainly in \(I_W\) in the W-dominated regime, in both sectors in the crossover regime, and primarily in \(I_X\) in the X-dominated regime. 
    (g)-(i) Corresponding linewidth. 
    The linewidth narrowing in a given sector provides a stronger indication of the formation of a polariton condensate.
}
    \label{fig:expe}
\end{figure}

Figure~\ref{fig:expe} summarizes the corresponding features that can be experimentally detected. The most direct probe is momentum-resolved photoluminescence. In the equilibrium-like regime Fig.~\ref{fig:expe}(a), the dominant emission hotspots appear at the W-valley positions, consistent with condensation at the lower-polariton minimum. The X points may still contribute a weak background, but they do not carry the macroscopic occupation. In the crossover regime Fig.~\ref{fig:expe}(b), both W and X sectors are populated. This situation is expected when the transfer time from X to W is comparable to the polariton lifetime, so that relaxation is neither fast enough to fully thermalize the distribution nor slow enough to isolate the X sector. In the strongly nonequilibrium regime Fig.~\ref{fig:expe}(c), the emission is dominated by the X region, which indicates that the high-density-of-states X sector acts as an efficient reservoir-fed accumulation channel.

The same distinction can be tested by measuring the pump-power dependence of the momentum-integrated emission from the two sectors. In the W-dominated case Fig.~\ref{fig:expe}(d), the W-sector intensity \(I_W\) shows the first nonlinear increase at the threshold, while \(I_X\) remains much weaker. This is the expected behavior when relaxation toward the true band minimum is efficient. In the crossover regime Fig.~\ref{fig:expe}(e), the two intensities increase at nearby pump powers, reflecting the coexistence of population in both regions. In contrast, in the X-dominated regime Fig.~\ref{fig:expe}(f), the first threshold-like enhancement occurs in \(I_X\).

Apart from the emission intensity, a high photoluminescence intensity may also arise from population accumulation in a relaxation bottleneck. For this reason, the linewidth provides a more stringent diagnostic. In the W-dominated regime Fig.~\ref{fig:expe}(g), the linewidth of the W-sector emission, \(\Gamma_W\), collapses near the threshold, while the X-sector linewidth changes only weakly. This behavior is consistent with coherent mode formation at the W valleys. In the crossover regime Fig.~\ref{fig:expe}(h), both linewidths narrow partially, reflecting the mixed character of the occupation. Finally, in the X-dominated regime Fig.~\ref{fig:expe}(i), the linewidth narrowing occurs first and most strongly in the X sector. Such a simultaneous observation of nonlinear intensity growth and linewidth collapse at X would support the formation of an X-dominated nonequilibrium coherent state.

\section{Conclusion}
\label{sec:conclusion}

In summary, we have developed a unified theory of three-dimensional photonic-crystal exciton-polaritons that treats symmetry-selected light-matter coupling, lower-polariton band geometry, equilibrium-like condensation, and driven-dissipative kinetics within a single framework. Our main advance is to show that, in a three-dimensional polaritonic band, the W-point global minimum and the nearby X-point saddle play fundamentally different but complementary roles: the former determines the true condensation geometry, while the latter reshapes the threshold and provides a competing kinetic accumulation channel. This establishes a band-structure-based picture of polariton condensation that goes beyond the standard planar-cavity paradigm.

These results identify three-dimensional photonic-crystal polaritons as a qualitatively new platform for condensation physics, where valleys, saddle structures, and inter-region relaxation pathways must be treated on equal footing. Looking forward, the present framework can be extended to multicomponent valley dynamics, more microscopic scattering processes, and interaction-driven nonlinear condensation phenomena, and it provides a natural starting point for quantitative comparison with future momentum-resolved and time-resolved experiments on three-dimensional polaritonic systems.

\begin{acknowledgments}
AVK acknowledges support from Saint Petersburg State University (Research Grant No. 125022803069-4) and from the Innovation Program for Quantum Science and Technology (No. 2021ZD0302704).
\end{acknowledgments}

\appendix
\section{Electric-field profiles and bright-mode selection}

To support the symmetry-based bright-band selection used in the main text, we show the Bloch electric field profiles of the target photonic modes. The plotted quantity is the field intensity $|\mathbf E_{n\mathbf k}(\mathbf r)|^2$ of the selected photonic branch. The field is concentrated near the internal dielectric-vacancy interface of the inverse-opal structure, where the excitonic medium is assumed to be distributed. This spatial overlap justifies the exciton-photon coupling model used in Eq.~\eqref{eq:g_overlap}, in which the effective coupling strength is determined by the overlap between the excitonic envelope and the Bloch electric field. In addition to the total field intensity, we plot the component-resolved Bloch electric fields $E_x$, $E_y$, and $E_z$ for the twofold-degenerate modes at the X- and W-points. These component-resolved field profiles provide a direct real-space diagnostic of how the degenerate partner modes transform under the corresponding little group operations. They therefore support the symmetry assignment of the selected bright photonic branch and justify the symmetry filtering used in constructing the effective exciton-photon Hamiltonian.

\begin{figure}
    \centering
    \includegraphics[width=1\linewidth]{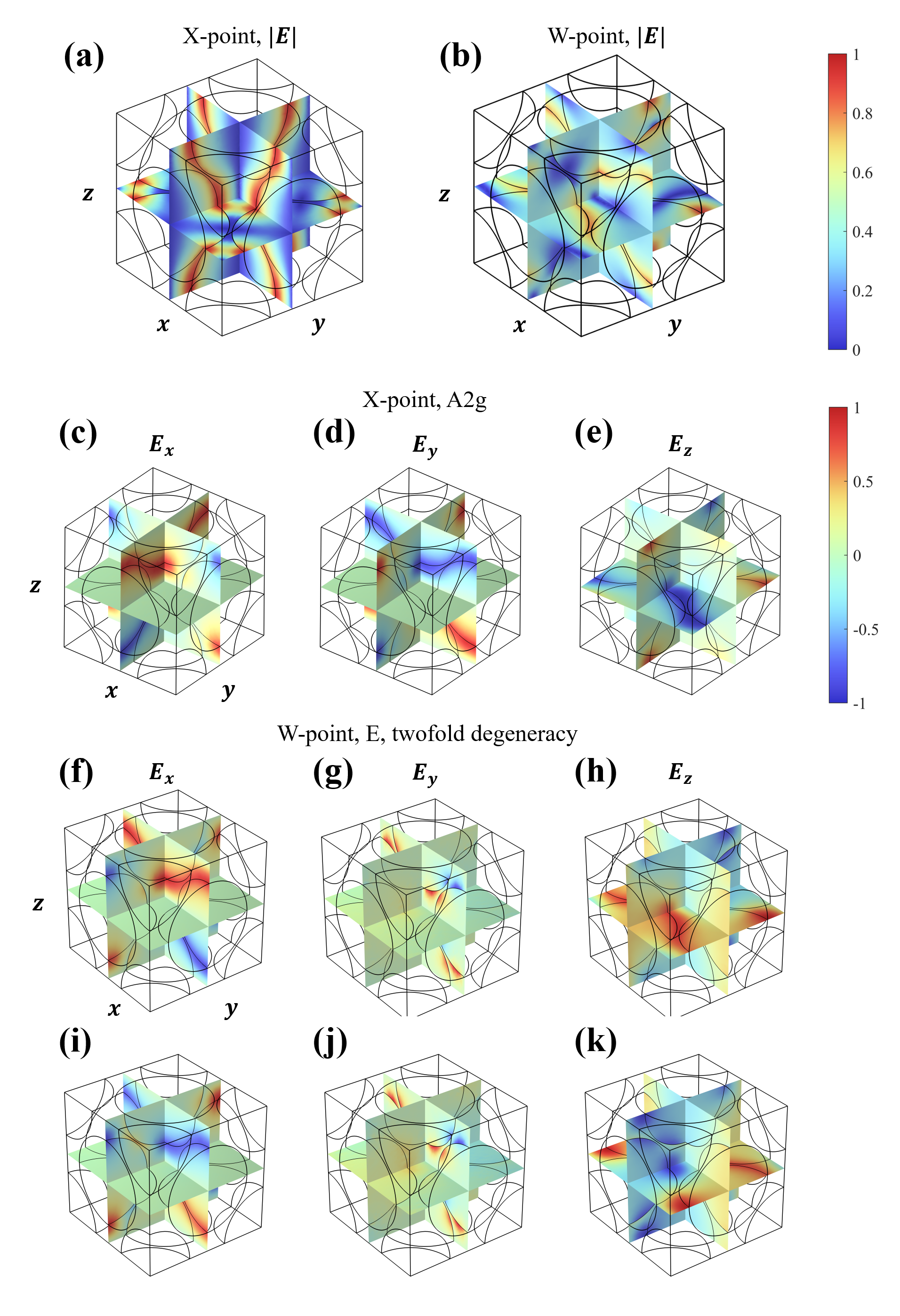}
    \caption{
Bloch electric-field profiles of the symmetry-selected photonic modes in the inverse-opal photonic crystal. (a-b) Field intensity $|\mathbf E_{n\mathbf k}(\mathbf r)|^2$ of the target bright branch near the W and X regions, respectively. The field is enhanced near the internal dielectric-vacancy interface, where the excitonic medium is assumed to be localized. These profiles provide a real-space visualization of the spatial-overlap criterion used to select the bright photonic branch entering the effective exciton-photon Hamiltonian. The component-resolved fields \(E_x\), \(E_y\), and \(E_z\) are shown for the twofold-degenerate modes at X (c-e) and W (f-k). Their transformation patterns under the corresponding little-group operations are used to identify the symmetry character of the selected bright photonic branch.
}
    \label{fig:Efield}
\end{figure}
\newpage

\bibliography{apssamp}

@book{kavokin2017microcavities,
  title={Microcavities},
  author={Kavokin, Alexey and Baumberg, Jeremy J and Malpuech, Guillaume and Laussy, Fabrice P},
  year={2017},
  publisher={Oxford university press}
}

@article{deng2010exciton,
  title={Exciton-polariton bose-einstein condensation},
  author={Deng, Hui and Haug, Hartmut and Yamamoto, Yoshihisa},
  journal={Reviews of modern physics},
  volume={82},
  number={2},
  pages={1489--1537},
  year={2010},
  publisher={APS}
}

@article{butov2012behaviour,
  title={The behaviour of exciton--polaritons},
  author={Butov, LV and Kavokin, AV},
  journal={Nature Photonics},
  volume={6},
  number={1},
  pages={2--2},
  year={2012},
  publisher={Nature Publishing Group UK London}
}

@article{cao2021strong,
  title={Strong light--matter coupling in microcavities characterised by Rabi-splittings comparable to the Bragg stop-band widths},
  author={Cao, Junhui and De Liberato, Simone and Kavokin, Alexey V},
  journal={New Journal of Physics},
  volume={23},
  number={11},
  pages={113015},
  year={2021},
  publisher={IOP Publishing}
}

@article{weisbuch1992observation,
  title={Observation of the coupled exciton-photon mode splitting in a semiconductor quantum microcavity},
  author={Weisbuch, Claude and Nishioka, Mr and Ishikawa, A and Arakawa, Y},
  journal={Physical review letters},
  volume={69},
  number={23},
  pages={3314},
  year={1992},
  publisher={APS}
}

@article{kasprzak2006bose,
  title={Bose--Einstein condensation of exciton polaritons},
  author={Kasprzak, Jacek and Richard, Murielle and Kundermann, Stefan and Baas, A and Jeambrun, P and Keeling, Jonathan Mark James and Marchetti, Francesca Maria and Szyma{\'n}ska, MH and Andr{\'e}, R and Staehli, JL a and others},
  journal={Nature},
  volume={443},
  number={7110},
  pages={409--414},
  year={2006},
  publisher={Nature Publishing Group UK London}
}

@article{wouters2008spatial,
  title={Spatial and spectral shape of inhomogeneous nonequilibrium exciton-polariton condensates},
  author={Wouters, Michiel and Carusotto, Iacopo and Ciuti, Cristiano},
  journal={Physical Review B—Condensed Matter and Materials Physics},
  volume={77},
  number={11},
  pages={115340},
  year={2008},
  publisher={APS}
}

@article{klembt2018exciton,
  title={Exciton-polariton topological insulator},
  author={Klembt, Sebastian and Harder, TH and Egorov, OA and Winkler, K and Ge, R and Bandres, MA and Emmerling, M and Worschech, L and Liew, TCH and Segev, M and others},
  journal={Nature},
  volume={562},
  number={7728},
  pages={552--556},
  year={2018},
  publisher={Nature Publishing Group UK London}
}

@article{hu2017imaging,
  title={Imaging exciton--polariton transport in MoSe2 waveguides},
  author={Hu, Fengrui and Luan, Yilong and Scott, ME and Yan, Jiaqiang and Mandrus, DG and Xu, Xiaodong and Fei, Z},
  journal={Nature Photonics},
  volume={11},
  number={6},
  pages={356--360},
  year={2017},
  publisher={Nature Publishing Group UK London}
}

@article{nitsche2014algebraic,
  title={Algebraic order and the Berezinskii-Kosterlitz-Thouless transition in an exciton-polariton gas},
  author={Nitsche, Wolfgang H and Kim, Na Young and Roumpos, Georgios and Schneider, Christian and Kamp, Martin and H{\"o}fling, Sven and Forchel, Alfred and Yamamoto, Yoshihisa},
  journal={Physical Review B},
  volume={90},
  number={20},
  pages={205430},
  year={2014},
  publisher={APS}
}

@article{imamog1996nonequilibrium,
  title={Nonequilibrium condensates and lasers without inversion: Exciton-polariton lasers},
  author={Imamog, A and Ram, RJ and Pau, S and Yamamoto, Y and others},
  journal={Physical Review A},
  volume={53},
  number={6},
  pages={4250},
  year={1996},
  publisher={APS}
}

@article{whittaker2018exciton,
  title={Exciton polaritons in a two-dimensional Lieb lattice with spin-orbit coupling},
  author={Whittaker, CE and Cancellieri, Emiliano and Walker, PM and Gulevich, DR and Schomerus, H and Vaitiekus, D and Royall, B and Whittaker, DM and Clarke, E and Iorsh, IV and others},
  journal={Physical review letters},
  volume={120},
  number={9},
  pages={097401},
  year={2018},
  publisher={APS}
}

@article{lin1998three,
  title={A three-dimensional photonic crystal operating at infrared wavelengths},
  author={Lin, Shawn-yu and Fleming, JG and Hetherington, DL and Smith, BK and Biswas, R and Ho, KM and Sigalas, MM and Zubrzycki, W and Kurtz, SR and Bur, Jim},
  journal={Nature},
  volume={394},
  number={6690},
  pages={251--253},
  year={1998},
  publisher={Nature Publishing Group UK London}
}

@article{moon2010chemical,
  title={Chemical aspects of three-dimensional photonic crystals},
  author={Moon, Jun Hyuk and Yang, Shu},
  journal={Chemical reviews},
  volume={110},
  number={1},
  pages={547--574},
  year={2010},
  publisher={ACS Publications}
}

@article{cersonsky2021diversity,
  title={The diversity of three-dimensional photonic crystals},
  author={Cersonsky, Rose K and Antonaglia, James and Dice, Bradley D and Glotzer, Sharon C},
  journal={Nature communications},
  volume={12},
  number={1},
  pages={2543},
  year={2021},
  publisher={Nature Publishing Group UK London}
}

@article{fleming2002all,
  title={All-metallic three-dimensional photonic crystals with a large infrared bandgap},
  author={Fleming, JG and Lin, SY and El-Kady, I and Biswas, R and Ho, KM},
  journal={Nature},
  volume={417},
  number={6884},
  pages={52--55},
  year={2002},
  publisher={Nature Publishing Group UK London}
}

@article{wouters2007excitations,
  title={Excitations in a nonequilibrium Bose-Einstein condensate of exciton polaritons},
  author={Wouters, Michiel and Carusotto, Iacopo},
  journal={Physical review letters},
  volume={99},
  number={14},
  pages={140402},
  year={2007},
  publisher={APS}
}

@book{pitaevskii2016bose,
  title={Bose-Einstein condensation and superfluidity},
  author={Pitaevskii, Lev and Stringari, Sandro},
  volume={164},
  year={2016},
  publisher={Oxford University Press}
}

@article{ibanescu2006enhanced,
  title={Enhanced photonic band-gap confinement via van hove saddle point singularities},
  author={Ibanescu, Mihai and Reed, Evan J and Joannopoulos, JD},
  journal={Physical review letters},
  volume={96},
  number={3},
  pages={033904},
  year={2006},
  publisher={APS}
}

@article{aoki2008coupling,
  title={Coupling of quantum-dot light emission with a three-dimensional photonic-crystal nanocavity},
  author={Aoki, Kanna and Guimard, Denis and Nishioka, Masao and Nomura, Masahiro and Iwamoto, Satoshi and Arakawa, Yasuhiko},
  journal={nature photonics},
  volume={2},
  number={11},
  pages={688--692},
  year={2008},
  publisher={Nature Publishing Group UK London}
}

@article{tandaechanurat2011lasing,
  title={Lasing oscillation in a three-dimensional photonic crystal nanocavity with a complete bandgap},
  author={Tandaechanurat, Aniwat and Ishida, Satomi and Guimard, Denis and Nomura, Masahiro and Iwamoto, Satoshi and Arakawa, Yasuhiko},
  journal={Nature Photonics},
  volume={5},
  number={2},
  pages={91--94},
  year={2011},
  publisher={Nature Publishing Group UK London}
}

@article{doosje2000photonic,
  title={Photonic bandgap optimization in inverted fcc photonic crystals},
  author={Doosje, Marcel and Hoenders, Bernhard J and Knoester, Jasper},
  journal={Journal of the Optical Society of America B},
  volume={17},
  number={4},
  pages={600--606},
  year={2000},
  publisher={Optical Society of America}
}

@article{tarhan1996photonic,
  title={Photonic band structure of fcc colloidal crystals},
  author={Tarhan, I Inanc and Watson, George H},
  journal={Physical review letters},
  volume={76},
  number={2},
  pages={315},
  year={1996},
  publisher={APS}
}

@article{miguez1998control,
  title={Control of the photonic crystal properties of fcc-packed submicrometer SiO2 spheres by sintering},
  author={M{\'\i}guez, Hern{\'a}n and Meseguer, Francisco and L{\'o}pez, Cefe and Blanco, {\'A}lvaro and Moya, Jos{\'e} S and Requena, Joaqu{\'\i}n and Mifsud, Amparo and Forn{\'e}s, Vicente},
  journal={Advanced Materials},
  volume={10},
  number={6},
  pages={480--483},
  year={1998},
  publisher={Wiley Online Library}
}

@article{cao2026emergent,
  title={Emergent Magnetic Monopole in Artificial Polariton Spin Ice},
  author={Cao, Junhui and Kavokin, Alexey},
  journal={arXiv preprint arXiv:2603.28384},
  year={2026}
}

@article{hopfield1958theory,
  title={Theory of the contribution of excitons to the complex dielectric constant of crystals},
  author={Hopfield, John J},
  journal={Physical Review},
  volume={112},
  number={5},
  pages={1555},
  year={1958},
  publisher={APS}
}

@article{gerace2007quantum,
  title={Quantum theory of exciton-photon coupling in photonic crystal slabs with embedded quantum wells},
  author={Gerace, Dario and Andreani, Lucio Claudio},
  journal={Physical Review B—Condensed Matter and Materials Physics},
  volume={75},
  number={23},
  pages={235325},
  year={2007},
  publisher={APS}
}

@article{zhang2018photonic,
  title={Photonic-crystal exciton-polaritons in monolayer semiconductors},
  author={Zhang, Long and Gogna, Rahul and Burg, Will and Tutuc, Emanuel and Deng, Hui},
  journal={Nature communications},
  volume={9},
  number={1},
  pages={713},
  year={2018},
  publisher={Nature Publishing Group UK London}
}

@article{whittaker2021exciton,
  title={Exciton--polaritons in GaAs-based slab waveguide photonic crystals},
  author={Whittaker, CE and Isoniemi, T and Lovett, S and Walker, PM and Kolodny, S and Kozin, V and Iorsh, IV and Farrer, I and Ritchie, DA and Skolnick, MS and others},
  journal={Applied Physics Letters},
  volume={119},
  number={18},
  year={2021},
  publisher={AIP Publishing}
}

@article{tetreault2004silicon,
  title={Silicon inverse opal—a platform for photonic bandgap research},
  author={Tetreault, N and M{\'\i}guez, Hern$\mu$n and Ozin, Geoffrey A},
  journal={Advanced Materials},
  volume={16},
  number={16},
  pages={1471--1476},
  year={2004},
  publisher={Wiley Online Library}
}

@article{waterhouse2007opal,
  title={Opal and inverse opal photonic crystals: Fabrication and characterization},
  author={Waterhouse, Geoffrey IN and Waterland, Mark R},
  journal={Polyhedron},
  volume={26},
  number={2},
  pages={356--368},
  year={2007},
  publisher={Elsevier}
}

@article{volovik2017topological,
  title={Topological lifshitz transitions},
  author={Volovik, GE},
  journal={Low Temperature Physics},
  volume={43},
  number={1},
  pages={47--55},
  year={2017},
  publisher={AIP Publishing}
}

@article{lifshitz1960anomalies,
  title={Anomalies of electron characteristics of a metal in the high pressure region},
  author={Lifshitz, IM and others},
  journal={Sov. Phys. JETP},
  volume={11},
  number={5},
  pages={1130--1135},
  year={1960}
}

@article{jalali2023topological,
  title={Topological Bogoliubov quasiparticles from Bose-Einstein condensate in a flat band system},
  author={Jalali-Mola, Zahra and Grass, Tobias and Kasper, Valentin and Lewenstein, Maciej and Bhattacharya, Utso},
  journal={Physical Review Letters},
  volume={131},
  number={22},
  pages={226601},
  year={2023},
  publisher={APS}
}

@article{reithmaier2004strong,
  title={Strong coupling in a single quantum dot--semiconductor microcavity system},
  author={Reithmaier, J Pelal and S{\k{e}}k, G and L{\"o}ffler, A and Hofmann, C and Kuhn, S and Reitzenstein, S and Keldysh, LV and Kulakovskii, VD and Reinecke, TL and Forchel, A},
  journal={Nature},
  volume={432},
  number={7014},
  pages={197--200},
  year={2004},
  publisher={Nature Publishing Group UK London}
}

@article{englund2010resonant,
  title={Resonant excitation of a quantum dot strongly coupled to a photonic crystal nanocavity},
  author={Englund, Dirk and Majumdar, Arka and Faraon, Andrei and Toishi, Mitsuru and Stoltz, Nick and Petroff, Pierre and Vu{\v{c}}kovi{\'c}, Jelena},
  journal={Physical Review Letters},
  volume={104},
  number={7},
  pages={073904},
  year={2010},
  publisher={APS}
}

@inproceedings{vuckovic2023scalable,
  title={Scalable semiconductor quantum systems},
  author={Vuckovic, Jelena},
  booktitle={Quantum Computing, Communication, and Simulation III},
  pages={PC124460H},
  year={2023},
  organization={SPIE}
}

@article{klimonsky2011photonic,
  title={Photonic crystals based on opals and inverse opals: synthesis and structural features},
  author={Klimonsky, Sergey Olegovich and Abramova, Vera V and Sinitskii, Alexander S and Tretyakov, Yuri D},
  journal={Russian Chemical Reviews},
  volume={80},
  number={12},
  pages={1191--1207},
  year={2011}
}

@article{richard2005experimental,
  title={Experimental evidence for nonequilibrium Bose condensation of exciton polaritons},
  author={Richard, Murielle and Kasprzak, J and Andr{\'e}, R and Romestain, R and Dang, Le Si and Malpuech, G and Kavokin, A},
  journal={Physical Review B—Condensed Matter and Materials Physics},
  volume={72},
  number={20},
  pages={201301},
  year={2005},
  publisher={APS}
}

\end{document}